PAPER

# Medical Support System for Spontaneous Breathing Trial Prediction Using Nonuniform Discrete Fourier Transform

Hernando González[1](✉),
Carlos Julio Arizmendi[1],
Beatriz F. Giraldo[2–4]

[1]Universidad Autónoma de Bucaramanga, Bucaramanga, Colombia

[2]Universitat Politècnica de Catalunya – Barcelona Tech (UPC), Barcelona, Spain

[3]Institute for Bioengineering of Catalonia (IBEC), The Barcelona Institute of Science and Technology, Barcelona, Spain

[4]CIBER de Bioingeniería, Biomateriales y Nanomedicina, Madrid, Spain

hgonzalez7@unab.edu.co

**ABSTRACT**

Spontaneous breathing trials (SBTs) represent a pivotal phase in the weaning process of mechanically ventilated patients. The objective of these trials is to assess patients' readiness to resume independent breathing, thereby facilitating timely weaning and reducing the duration of mechanical ventilation (MV). Nevertheless, accurately predicting the success or failure of SBT remains a significant challenge in clinical practice. This study proposes a healthcare system that employs machine learning techniques to predict the outcome of SBT. The model is trained on respiratory flow and electrocardiogram (ECG) signals, employing the non-uniform discrete Fourier transform (NUDFT) for frequency domain analysis. The SBT prediction model has the potential to significantly enhance clinical decision-making by enabling the early identification of patients at risk for SBT failure, achieving an accuracy of 84.4 ± 3.2%.

## 1    INTRODUCTION

Mechanical ventilation (MV) is a cornerstone of critical care medicine, providing essential respiratory support to patients with acute respiratory failure or conditions that impair their ability to breathe adequately [1, 2]. However, prolonged use of MV can lead to several complications, including ventilator-associated pneumonia, barotrauma, and muscle weakness, which together contribute to increased morbidity and mortality rates [3, 4]. Thus, achieving timely liberation from MV through successful weaning is crucial for optimizing patient outcomes and reducing healthcare costs [5].

The spontaneous breathing trial (SBT) is a critical step in the weaning process, as it assesses a patient's readiness to breathe independently without ventilator support. During an SBT, patients are allowed to breathe spontaneously while still connected to the ventilator, enabling clinicians to evaluate their ability to maintain adequate

González, H., Arizmendi, C.J., Giraldo, B.F. (2024). Medical Support System for Spontaneous Breathing Trial Prediction Using Nonuniform Discrete Fourier Transform. *International Journal of Online and Biomedical Engineering (iJOE)*, 20(16), pp. 103–116. https://doi.org/10.3991/ijoe.v20i16.52859

Article submitted 2024-09-01. Revision uploaded 2024-10-14. Final acceptance 2024-10-17.

gas exchange and respiratory function. Successful completion of an SBT suggests that the patient can be weaned, facilitating the transition to independent breathing. Predicting SBT outcomes despite their clinical importance remains a significant challenge. Clinicians typically rely on subjective assessments and predefined criteria, which may lack sensitivity and specificity, potentially leading to delays in weaning or premature weaning, both of which can negatively affect patient outcomes [6, 7].

The non-uniform discrete Fourier transform (NUDFT) has emerged as a valuable tool in digital signal processing, offering higher resolution compared to the traditional discrete Fourier transform (DFT) by sampling data at non-uniform intervals in the time or frequency domain [8]. This technique has been successfully applied in various fields, including spectral domain optical coherence tomography (SD-OCT), where it improves sensitivity and reduces processing time [9]. Additionally, NUDFT has shown advantages in handling under sampled data, mitigating aliasing and noise, and reducing memory consumption [10, 11].

This study aims to leverage the power spectral density (PSD) features extracted from respiratory flow and electrocardiographic (ECG) signals to develop a machine learning classifier that predicts a patient's likelihood of successful weaning after an SBT. By utilizing NUDFT to analyze the irregular temporal patterns in the physiological data collected during SBTs, this approach offers a more precise tool for evaluating patient readiness. Given the critical role of SBTs in the weaning process, accurate predictions of their success are essential not only to reduce the duration of MV but also to minimize complications such as ventilator-associated infections and muscle atrophy. Failed weaning attempts, often resulting from premature extubation, can lead to respiratory distress, reintubation, and prolonged intensive care unit (ICU) stays, increasing morbidity and healthcare costs. Existing prediction methods, based largely on subjective judgment and basic physiological thresholds, frequently prove inadequate. This study aims to address these limitations by applying machine learning techniques and the NUDFT to respiratory and ECG data, offering a novel, data-driven approach to enhance the accuracy of SBT outcome predictions. By improving the reliability of these predictions, this study seeks to contribute to safer and more efficient weaning practices in critical care settings.

## 2 MATERIALS AND METHODS

### 2.1 Database

The Weandb database is derived from a study that encompassed ECG and respiratory flow signals from 133 mechanically ventilated patients who underwent weaning [12]. For each patient, eight time series were recorded, including the RR interval between consecutive beats of an ECG signal, inspiration time ($T_I$), expiration time ($T_E$), respiratory cycle duration ($T_{Tot} = T_I + T_E$), tidal volume ($V_T$), inspiratory fraction ($T_I/T_{Tot}$), mean inspired flow ($V_T/T_I$), and frequency-tidal volume ratio ($f/V_T$). These parameters, derived from the respiratory flow signal and ECG, are measured at specific time points, resulting in a non-uniform time vector. This non-uniformity reflects the variability in the intervals between consecutive measurements. The database comprises three distinct classes:

– The success group (Class $C_0$) comprises 94 patients who successfully completed the weaning process.
– The failure group (Class $C_1$) encompasses 39 patients who failed to sustain spontaneous breathing and required reconnection to the ventilator within 30 minutes.
– The reintubated group (Class $C_2$) includes 21 patients who initially passed the 30-minute SBT test but required reintubation and were reconnected to MV within the next 48 hours.

Classes $C_1$ and $C_2$ were merged into a single failure category, as both groups experienced unsuccessful spontaneous breathing followed by reintubation or reconnection to MV. This consolidation enhances the dataset's clarity and relevance for distinguishing between successful and unsuccessful leaning outcomes. A data-wrangling methodology was implemented to eliminate outliers and identify the most continuous information region. Initially, all null values were removed from the dataset. Outliers were identified by comparing the absolute difference between each data point and the mean value, normalized by the standard deviation, against a predefined threshold. Outliers were then replaced by the mean of the surrounding data points, calculated based on neighboring values within a specified range. During the recording of the respiratory flow signal time series, episodes of apnea were observed. Therefore, the signal was reviewed, and the longest continuous time interval was registered for analysis. Subsequently, uninterrupted data regions were extracted by identifying gaps where the time difference between consecutive points exceeded a certain threshold. Each segment of consistent data was stored separately, and the longest uninterrupted segment was selected for further analysis. Subsequently, the signal data were normalized by centering them around zero (subtracting the mean) and scaling to unit variance (dividing by the standard deviation). This normalization was performed to facilitate comparison and analysis of the signals at different patients and time points. Figure 1 shows the time series after applying the normalization and data processing procedures for a patient in the success group.

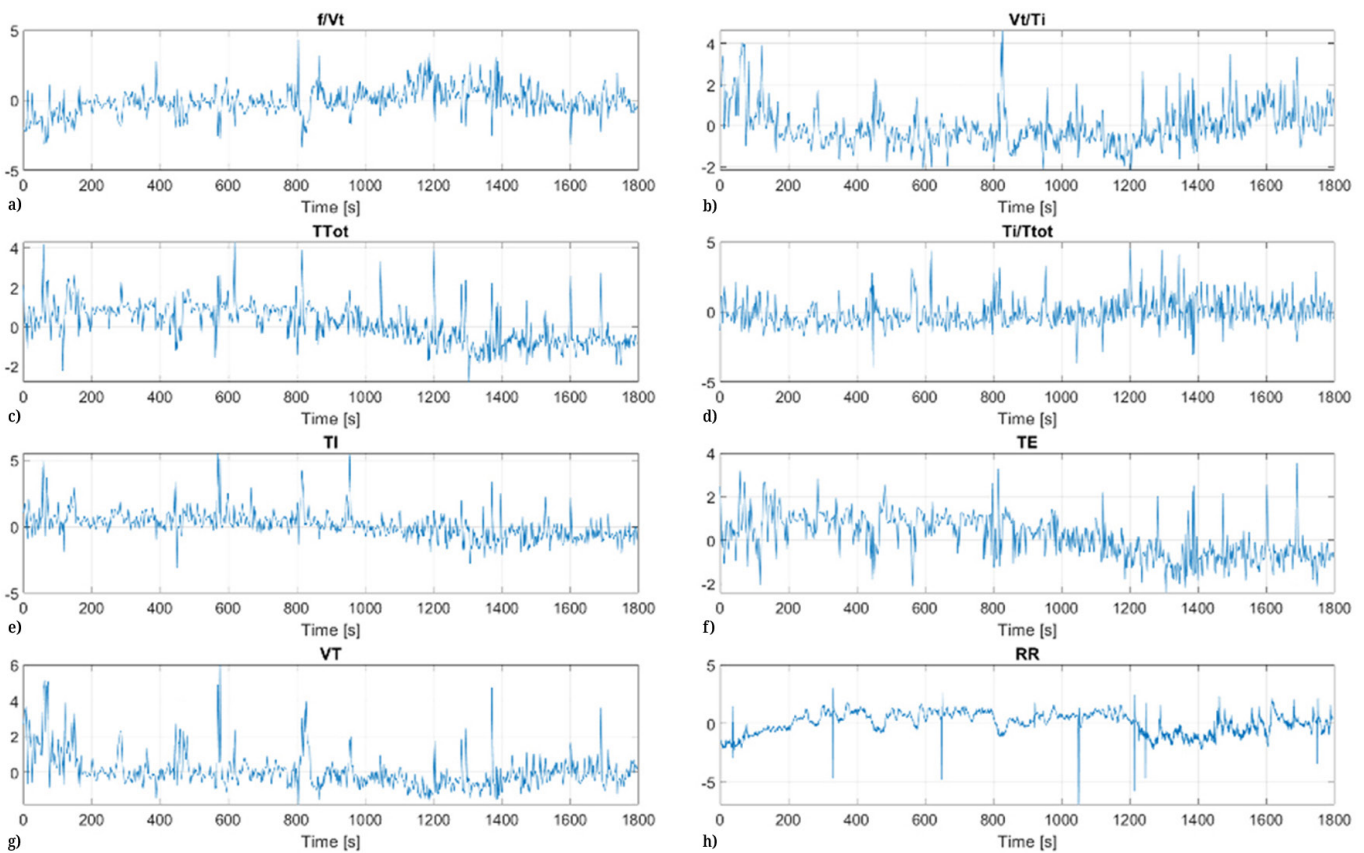

**Fig. 1.** An example of the different time series of a patient in the success group, after preprocessing the signals. a) Frequency-Tidal volume ratio $f/V_T$, b) Mean inspired flow $V_T/T_I$, c) Breathing duration $T_{Tot}$, d) Inspiratory fraction $T_I/T_{Tot}$, e) Inspiratory time $T_I$, f) Expiratory time $T_E$, g) Tidal volume $V_T$, h) Beat to beat interval $RR$

### 2.2 Non-uniform discrete Fourier transform

The NUDFT is the non-uniform variant of the regular DFT. The NUDFT is defined as

$$F(v_k) = \sum_{n=1}^{N} f(x_n)e^{-i2\pi v_k x_n} \quad k = 1,2,\ldots M \quad (1)$$

where, $v_k$ represents frequency in the $k_{th}$ spectral point, $x_n$ is the position of the $n_{th}$ sampling point, $f(x_n)$ is the magnitude of the signal at the sampled position, and $F(v_k)$ is the spectral intensity at the frequency.

Interpolation of signals can introduce discrepancies in the PSD of non-uniformly sampled signals, as shown in Figure 2, which compares the PSD of the original $f/V_t$ signal from the success group with the PSD obtained using two interpolation methods: linear and cubic spline. Although the overall shape of the PSD remains consistent across the interpolated signals, significant differences are observed when compared to the PSD derived from the NUDFT. Specifically, in the low-frequency range, the amplitude of the interpolated signals is lower than that of the original PSD. This reduction in amplitude is likely due to interpolation, which can smooth out variations in the data and attenuate lower frequencies because of the approximation between sampling points. Furthermore, a pronounced peak in the PSD at the sampling frequency appears in the interpolated signals, which is absent in the NUDFT-derived PSD. This peak is likely a result of interpolation artifacts, which are unintended distortions introduced during the process, manifesting as spurious components within the signal. These artifacts typically manifest as false frequencies or exaggerated spectral content, particularly when the signal has sharp transitions or rapid variations. Such effects are well-documented in spectral analysis, where interpolation, especially when applied to non-uniformly sampled data, can inadvertently modify the true spectral characteristics of the signal. Therefore, due to the distortions introduced by interpolation, in this study, the time-frequency plot will be generated using the NUDFT to ensure the preservation of spectral integrity. While popular time-frequency representations such as the continuous wavelet transform (CWT) and the short-time Fourier transform (STFT) are often used, both require prior interpolation of the data, which can compromise the accuracy of the spectral information.

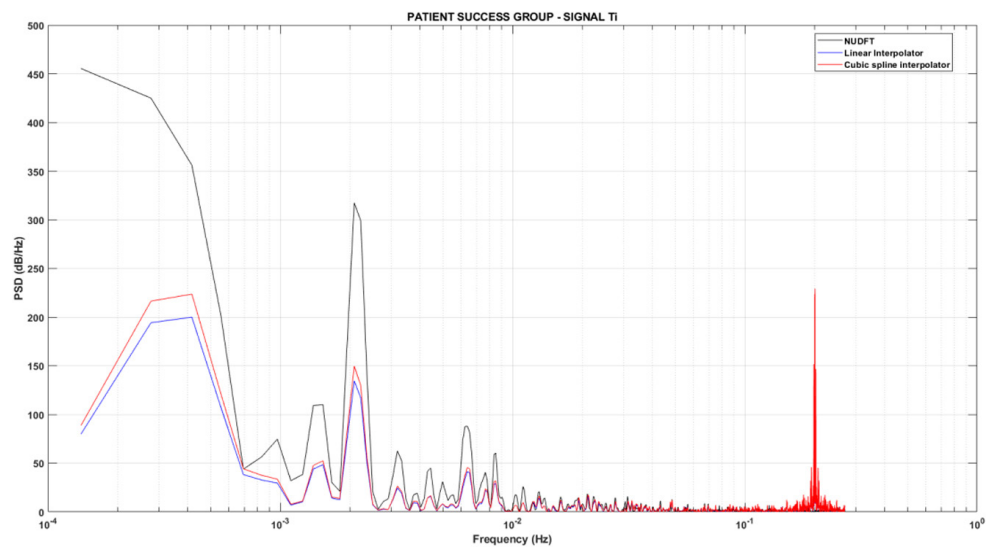

**Fig. 2.** Comparison of power spectral density of the $f/V_t$ signal for a patient in the success group: NUDFT vs. interpolated signals using the linear and cubic spline methods

Medical Support System for Spontaneous Breathing Trial Prediction Using Nonuniform Discrete Fourier Transform

Time-frequency representations were used to analyze the spectral evolution of the signals over time. A Hamming window of 100 seconds with 75% overlap was applied to each signal. This 75% overlap improves the temporal resolution by increasing the frequency of the sampling intervals, allowing more accurate detection of changes over time. The use of an hamming window, with its reduced spectral leakage properties and emphasis on resolution, is particularly well suited to capture the nuanced temporal features essential for monitoring respiratory and cardiac dynamics during weaning. The ability of the hamming window to minimize sidelobe levels in the frequency spectrum ensures that signal energy is more concentrated around its true frequency components, thus providing a clearer and more accurate representation of physiological variations within the clinically relevant time frame. Figure 3 shows the distribution of signal power across different frequency components over time, providing information on the dynamic changes in respiratory and cardiac activities of a patient in the success group for the eight-time series.

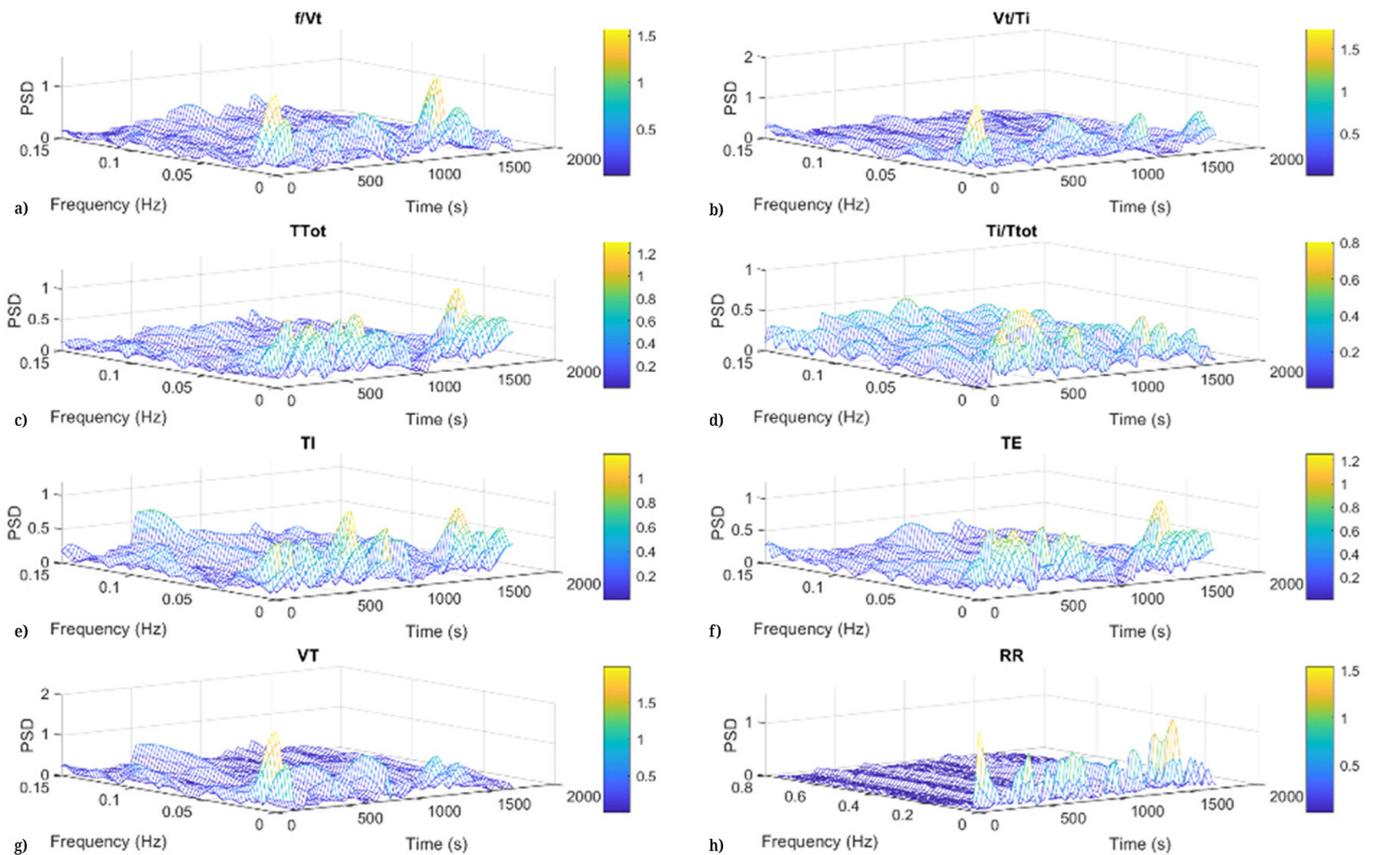

**Fig. 3.** Power spectral density as a function of time and frequency for a patient in the success group. a) Frequency-Tidal volume ratio $f/V_T$, b) Mean inspired flow $V_T/T_P$ c) Breathing duration $T_{Tot}$, d) Inspiratory fraction $T_I/T_{Tot}$, e) Inspiratory time $T_I$, f) Expiratory time $T_E$, g) Tidal volume $V_T$, h) Beat to beat interval $RR$

### 2.3 Features in frequency domain

For each 100-second time interval, the NUDFT was applied to the signal, followed by the computation of the PSD. The resulting time-frequency plot, like the STFT but utilizing the NUDFT in place of the FFT, offers a detailed representation of the signal's spectral content over time. From each time segment, various spectral features are extracted, providing valuable insights into the frequency components and enabling a more accurate characterization of the signal's temporal dynamics.

**Instantaneous frequency (IF).** This metric reflects the frequency content of the signal at specific time points within the window. It provides information about dynamic changes in frequency over time, giving a detailed view of how the spectral characteristics of the signal evolve [13–14].

**Mean frequency (MNF).** The MNF represents the MNF weighted by the PSD. It serves as a central measure of the frequency distribution within the signal, providing summary statistics of its dominant frequency components [15].

**Median frequency (MDF).** The MDF is the frequency value that separates the power spectrum into two equal areas. It provides a robust measure of the central tendency of the signal in terms of frequency distribution and is less sensitive to outliers compared to the mean frequency.

**Spectral entropy (SE).** SE quantifies the complexity or randomness of the spectral content. It measures the degree of irregularity or unpredictability of the frequency distribution, with higher values indicating greater diversity or complexity of frequency components [16].

**Spectral energy (SEn).** SEn is calculated as the sum of the squared magnitudes of the spectral components. This metric provides a measure of the signal's overall power and is useful in comparing the energy levels across different signals or time intervals [17].

**Spectral contrast (SC).** SC measures the difference in amplitude between the peaks and valleys in the spectral bands. It is computed by dividing the spectrum into several bands and calculating the difference between the maximum and minimum magnitudes within each band. This parameter helps in identifying the presence of harmonic structures and the distribution of energy across the spectrum.

**Spectral flatness (SF).** SF is the ratio of the geometric mean to the arithmetic mean of the power spectrum. It quantifies how flat or peaked the spectrum is, with higher values indicating a flatter spectrum. This metric is useful for distinguishing between tonal and noise-like signals, as tonal signals tend to have lower SF while noise-like signals have higher values.

**Spectral crest factor (SCF).** SCF is the ratio of the peak magnitude to the root mean square (RMS) magnitude of the spectrum. It indicates the presence of peaks within the spectrum, with higher values suggesting more pronounced peaks. This parameter is beneficial for identifying transient or impulsive components within the signal.

Figure 4 shows the frequency-domain features extracted from the $f/V_t$ signal for one patient from the success group and one from the failure group. The temporal variability across all descriptors underscores the importance of frequency-domain analysis for understanding respiratory signal dynamics and differentiating between patients with different clinical outcomes. Such signal processing techniques are crucial for identifying spectral patterns that may be indicative of a patient's respiratory status, offering potential markers for prognosis and clinical intervention. Based on the frequency features described for each interval, the following statistical descriptors are calculated to provide a comprehensive analysis of the signal properties: mean (*M*), standard deviation (*Std*), interquartile range (*Iq*), skewness (*S*), kurtosis (*K*), median (*Me*), and root mean square (*RMS*). A total of 448 features are extracted for each patient, based on seven respiratory flow signals and the *RR* signal, with each signal analyzed across eight spectral characteristics and seven statistical measures. To reduce the dimensionality of the system, the Mann-Whitney test is used to identify features with significant differences between groups, selecting those with a *p* value < 0.05. This nonparametric test evaluates whether two independent samples come from the same distribution, which makes it suitable for comparing continuous or ordinal data without assumptions of normality. The Mann-Whitney test consists of jointly ranking all observations in both groups and comparing the sum of the rankings of each group to assess whether the distributions differ significantly [18]. It is especially effective for data with non-normal distribution or cases where the assumptions of parametric tests are violated. Table 1 presents the 18 most relevant features selected for the classification system design, based on the *p*-value, along with their mean and standard deviation.

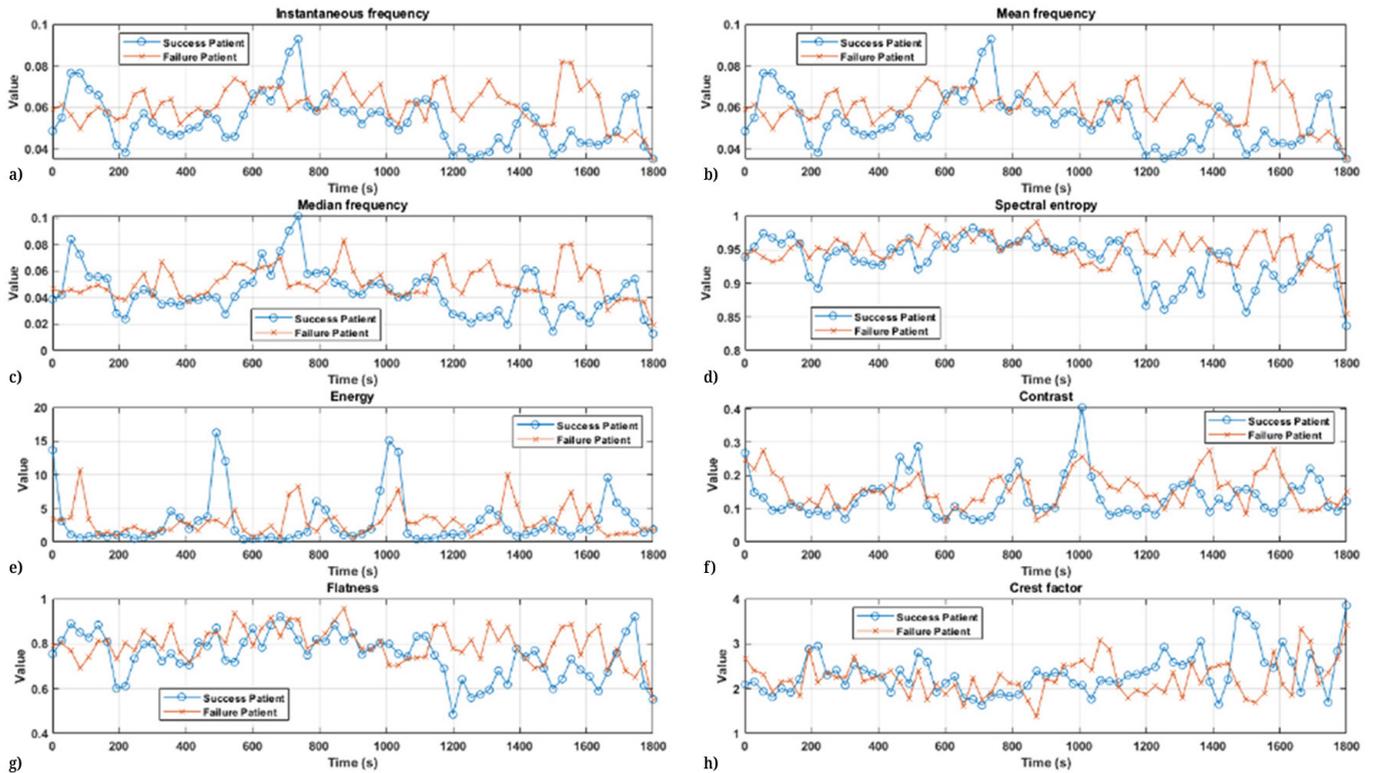

**Fig. 4.** Frequency domain features extracted from the $f/V_t$ signal. a) instantaneous frequency, b) mean frequency, c) median frequency, d) spectral entropy, e) energy, f) contrast, g) flatness, h) crest factor

Table 1. Feature selection summary with associated *p*-values

| Feature | Success Group | Failure Group | p-Value |
|---|---|---|---|
| $RMS[SEn(f/V_T)]$ | 3.1978+/−0.79631 | 2.8025+/−0.85856 | 0.001488 |
| $M[SC(f/V_T)]$ | 2.8818+/−0.70932 | 2.5558+/−0.92075 | 0.0028075 |
| $Std[SC(f/V_T)]$ | 3.1001+/−0.71109 | 2.7687+/−0.84316 | 0.0075979 |
| $Iq[SC(f/V_T)]$ | 3.1533+/−0.7753 | 2.7436+/−0.69106 | 0.00015166 |
| $Std[SF(f/V_T)]$ | 0.15449+/−0.021901 | 0.14346+/−0.02091 | 0.0021704 |
| $S[SF(f/V_T)]$ | 0.15752+/−0.023798 | 0.14723+/−0.025574 | 0.001546 |
| $Iq[SF(V_T/T_I)]$ | 3.9795+/−2.9673 | 2.8129+/−1.5528 | 0.0022804 |
| $Std[SCF(T_{Tot})]$ | 2.3755+/−0.91346 | 2.0102+/−0.96901 | 0.0046192 |
| $Iq[SCF(T_{Tot})]$ | 2.1977+/−0.84968 | 1.923+/−1.0531 | 0.0041111 |
| $K[SCF(T_{Tot})]$ | 2.4077+/−0.84053 | 2.061+/−1.0029 | 0.00084892 |

*(Continued)*

Table 1. Feature selection summary with associated *p* values *(Continued)*

| Feature | Success Group | Failure Group | p-Value |
|---|---|---|---|
| $Iq[IF(T_I/T_{Tot})]$ | 0.07456+/–0.022193 | 0.067658+/–0.02414 | 0.0051835 |
| $Me[MDF(V_T)]$ | 2.0443+/–0.72764 | 1.7788+/–0.75152 | 0.0058755 |
| $M[SE(V_T)]$ | 2.1856+/–0.72997 | 1.8544+/–0.78732 | 0.0021172 |
| $RMS[SE(V_T)]$ | 0.1419+/–0.027855 | 0.13046+/–0.023926 | 0.0054261 |
| $Me[SE(RR)]$ | 4.9314+/–2.2619 | 4.1991+/–1.8971 | 0.0040158 |
| $M[SEn(RR)]$ | 5.1656+/–2.7689 | 4.0335+/–1.4901 | 0.0011054 |
| $M[SC(RR)]$ | 0.16797+/–0.022778 | 0.15646+/–0.023096 | 0.0055512 |
| $Iq[SC(RR)]$ | 0.17202+/–0.026169 | 0.1636+/–0.026325 | 0.0064299 |

*Notes:* $M$ = mean, $Std$ = standard deviation, $Iq$ = interquartile range, $S$ = skewness, $K$ = kurtosis, $Me$ = median, $RMS$ = root mean square, $IF$ = instantaneous frequency, $MNF$ = mean frequency, $MDF$ = median frequency, $SE$ = spectral entropy, $SEn$ = spectral energy, $SC$ = spectral contrast, $SF$ = spectral flatness, $SCF$ = spectral crest factor, $T_I$ = inspiratory time, $T_E$ = expiratory time, $T_{Tot}$ = respiratory cycle duration, $V_T$ = tidal volume, $T_I/T_{Tot}$ = inspiratory fraction, $V_T/T_I$ mean inspired flow, $f/V_T$ frequency-tidal volume ratio, $RR$ interval between consecutive beats.

## 2.4 Classification system

The classification system is designed using support vector machines (SVM). This supervised learning algorithm searches for the optimal hyperplane that best separates data points into different classes in a high-dimensional space. SVM works by identifying support vectors-data points that are closest to the decision boundary-and maximizing the margin (distance) between the support vectors and the hyperplane, which helps improve the generalizability of the model [19–20]. The Bayesian optimization algorithm was used to optimize the parameters of the SVM algorithm [21–22]. This method systematically searches the parameter space by constructing a probabilistic model of the objective function and iteratively selecting the points to be evaluated based on this model. Prior to model training, the dataset was preprocessed by normalizing each feature to ensure all variables are on a comparable scale. The dataset was partitioned, with 70% allocated for training and 30% for testing. To address the class imbalance, class weights were adjusted within the SVM optimization framework.

The optimization process aims to maximize the Area under the Receiver Operating Characteristic Curve (AUC-ROC) by adjusting critical parameters: the kernel function (linear, polynomial, or radial basis function (RBF)), the regularization parameter (C), the polynomial order, and the kernel scale. The AUC-ROC is chosen as the optimization criterion due to its advantages in evaluating model performance, particularly in the presence of class imbalance. The AUC-ROC measures the ability of the model to distinguish between classes and is calculated as the integral of the ROC curve, which plots the true positive rate (sensitivity) against the false positive rate (1-specificity). By focusing on the AUC-ROC, the optimization process ensures a robust evaluation of the model's discriminatory power, leading to better generalization and performance across various thresholds. The objective function is defined to maximize the mean AUC-ROC between the training and test datasets, incorporating class weights to address class imbalance effectively. The formula for calculating the AUC-ROC is

$$AUC = \int_0^1 TPR(FPR)d(FPR) \qquad (2)$$

where, TPR is the true positive rate and FPR is the false positive rate. The algorithms were implemented in Python 3.10 using the *scikit-learn* library, with hyperparameter optimization carried out through the Bayesian optimization algorithm, implemented via the *bayes_opt* package. The optimization process was executed over 100 iterations, beginning with an initial random search of 10 samples to initialize the model. The best model obtained from the Bayesian optimization algorithm comprises an RBF kernel with a regularization parameter C of 0.25036 and a kernel scale of 2.828. The RBF kernel is characterized by its ability to handle nonlinear relationships by mapping input features into higher dimensional spaces. The regularization parameter C controls the tradeoff between achieving low training error and minimizing model complexity, thus avoiding overfitting. The kernel scaling parameter determines the width of the RBF kernel, which influences the flexibility of the decision boundary in the feature space.

## 3   RESULTS

Once the architecture is selected, the model is evaluated 150 times using four-fold cross-validation. The model is trained on three of the folds and validated on the remaining fold in each iteration. This process is repeated four times, with each fold used exactly once as a validation set. By averaging the performance metrics across these iterations, four-fold cross-validation provides a robust assessment of the model's ability to generalize and helps mitigate problems of overfitting and underfitting. This approach ensures consistent performance across dataset subsets, providing reliable estimates for real-world applications.

Table 2 shows the metrics for validation data: accuracy, precision, recall, F1 score, specificity, and the area under the AUC-ROC. These metrics provide a comprehensive assessment of the model's predictive performance and its effectiveness in classifying instances across different datasets. Precision measures the accuracy of positive predictions, while recall evaluates the model's ability to correctly identify positive instances. The F1 score, which is the harmonic mean of precision and recall, offers a balanced evaluation by considering both false positives and false negatives. Specificity measures the model's ability to correctly identify negative cases. The AUC indicates the model's capability to distinguish between positive and negative classes, with a higher AUC representing better overall performance. The results indicate that the classifier demonstrates consistent performance on both training and validation data sets, also indicating the model's balanced ability to correctly identify positive cases and its consistent performance in predicting positive cases.

Table 2. Performance metrics of classifier on training and validation data

| Metric | Validation Dataset |
| --- | --- |
| Accuracy | 84.4 ± 3.2% |
| Precision | 85.7 ± 4.5% |
| Recall | 87.5 ± 2.5% |
| F1 Score | 86.6 ± 3.4% |
| Specificity | 85.1 ± 4.1% |
| AUC | 84.8 ± 2.4% |

Despite the promising results, certain limitations of the proposed SVM-based classifier should be acknowledged. Although Bayesian optimization was used to efficiently search the hyperparameter space and identify the optimal values for the regularization parameter (C) and kernel type, the model's performance is still constrained by the characteristics of the dataset. Specifically, the relatively small dataset, particularly in the failure group, limits the generalization potential of the classifier. While 4-fold cross-validation provides a robust estimate of the model's ability to generalize, the dataset's size, especially in the failure group, may introduce biases when evaluating minority classes and affect the classifier's performance on unseen data in real clinical settings. Future work would benefit from increasing the dataset size, particularly by collecting more data from patients in the failure group or employing data augmentation techniques to synthetically expand the dataset. In addition, the SVM classifier used in this study assumes a linear or nonlinear boundary defined by a fixed kernel. Although the RBF kernel performed well, other machine learning models, such as neural networks or ensemble methods (e.g., random forests), may better capture complex patterns in the data and further improve predictive performance.

## 4 DISCUSSION

The literature review has highlighted various models developed from machine learning algorithms to predict the success or failure of MV weaning based on retrospectively collected data. These studies have employed methodologies such as neural networks, SVM, logistic regression, and deep learning, applied to both adult and pediatric populations. For instance, in the pediatric domain, [23] utilized machine learning techniques to predict weaning outcomes, demonstrating the applicability of advanced methods across different age groups. The dataset for this study includes the same variables referenced in these studies, along with additional variables such as ventilatory rate, peak inspiratory pressure, positive end-expiratory pressure, respiratory rate, and transcutaneous oxygen ($O_2$) [24–26]. A key distinction of this work lies in the implementation of the NUDFT for the frequency-domain analysis of respiratory flow and ECG signals during SBTs. Unlike traditional Fourier transform methods, which assume uniform sampling intervals, the NUDFT accommodates irregularly sampled data—an often encountered challenge in clinical environments. This allows for a more accurate frequency representation of physiological signals and makes NUDFT comparable to the STFFT when dealing with non-uniformly sampled data.

This study presents a methodology for time-frequency analysis in the medical domain, capturing subtle signal variations that are crucial for evaluating patient readiness for weaning. The results of this study demonstrate that the application of NUDFT, combined with machine learning classifiers based on features extracted from the PSD of respiratory and ECG signals, allows for the accurate prediction of weaning success. The model metrics, including accuracy, recall, and AUC, consistently show robust and reliable performance in classifying SBT outcomes. These results indicate that the proposed methodology not only improves prediction accuracy over traditional methods but also offers a more sophisticated tool for assessing patient readiness. The approach addresses limitations in existing methods, which heavily rely on subjective clinical judgment and simple physiological thresholds.

This methodology's implementation in clinical settings offers significant advantages for decision-making in ICUs. By improving diagnostic precision, this approach enables continuous patient monitoring and facilitates timely interventions. Additionally, advances in mobile health applications for the diagnosis and management of respiratory diseases complement the frequency-domain analysis performed in the ICU, extending the reach of diagnostic tools to remote settings and improving patient care beyond the hospital environment [27, 28].

## 5 CONCLUSION

The application of frequency analysis on respiratory and ECG signals from patients undergoing SBT using NUDFT offers a distinct advantage over traditional Fourier transform methods by accommodating irregularly sampled medical data. This adaptation preserves essential signal characteristics, ensuring greater accuracy and reliability in predicting SBT outcomes, ultimately enhancing patient care and clinical decision-making in ICUs. The integration of advanced signal processing techniques, such as NUDFT, with established clinical protocols facilitates more precise monitoring and informed decisions in the ICU. Additionally, incorporating machine learning into this framework holds significant potential to improve diagnostic and therapeutic capabilities in critical care, leading to better patient outcomes.

The proposed methodology introduces a novel, precise tool for analyzing non-uniformly sampled physiological data, which, unlike STFFT, is tailored to real-world clinical conditions. The results demonstrate that this approach significantly enhances the ability to capture relevant time-frequency patterns, improving the accuracy of SBT outcome predictions. The combination of advanced signal processing and machine learning contributes to safer, more efficient weaning practices in critical care.

## 7 AUTHORS

**Hernando González** received the B.Eng. degree in electronic engineering from Universidad Industrial de Santander (UIS), Master in Electronic Engineering from UIS, student of Doctorate in Engineering from UNAB, currently is Associate Professor of the Mechatronics Engineering program at UNAB and member of GICITII, His research interests include Machine Learning, Control Systems, signal treatments, robotics, electronics and mechatronics devices (E-mail: hgonzalez7@unab.edu.co).

**Carlos Julio Arizmendi** received the B.Eng. degree in electronic engineering from Universidad Industrial de Santander (UIS), Bucaramanga, Colombia in 1997, in 2008 received the diploma of advanced studies in the doctorate of biomedical engineering from UPC, in 2012 the PhD in Artificial Intelligent from UPC. He is currently Titular Professor of the Mechatronics Engineering program and Biomedical Engineering program at Universidad Autonoma de Bucaramanga (UNAB), also is the Director of the GICITII research group at UNAB, creator of the Biomedical Engineering Program at UNAB. His research interests include machine learning, deep learning, signal treatments, electronics, mechatronics and biomedical devices (E-mail: carizmendi@unab.edu.co).

**Beatriz F. Giraldo** received the B.Eng. degree in electrical engineering from the Technical University of Pereira, Pereira, Colombia, in 1983, the M.Sc. degree in Bioengineering, and the Ph.D. degree in Industrial Engineering in the Biomedical

Engineering Program from the Technical University of Catalonia (UPC), Barcelona, Spain, in 1990 and 1996, respectively. She is Associate Professor at the Automatic Control Department, UPC, and senior researcher of the Biomedical Signal Processing and Interpretation (BIOSPIN) group of the Institute for Bioengineering of Catalonia (IBEC), and the CIBER de Bioingeniería, Biomateriales y Nanomedicina (CIBER-BBN), in Spain. Her main research interests include biomedical signal processing and statistical analysis of cardiac, respiratory, and cardiorespiratory signals. Current research projects include study of cardiac and respiratory system in elderly patients, chronic heart failure patients, and patients on weaning trial process (E-mail: Beatriz.Giraldo@upc.edu, bgiraldo@ibecbarcelona.eu).